\documentclass[a4paper]{article}
\usepackage{epsfig,amsmath,amsfonts,amssymb,setspace,multirow,textcomp, subcaption}
\usepackage[T1]{fontenc}
\usepackage[sort&compress]{natbib}
\makeatletter
\renewcommand{\@oddhead}{\textit{} \hfil}
\renewcommand{\@evenfoot}{\hfil \thepage \hfil}
\renewcommand{\@oddfoot}{\hfil \thepage \hfil}

\newcommand{\grl}{    {\it Geophys. Res. Lett.}}

\newcommand{\jgr}{    {\it J. Geophys. Res.}}

\newcommand{\nat}{    {\it Nature}}

\newcommand{\planss}{ {\it Plan. Space Science}}

\newcommand{\ssr}{    {\it Space Sci. Rev.}}

\makeatother

\renewenvironment{thebibliography}[1]{\begin{oldthebibliography}{#1}\setlength{\parskip}{0ex}\setlength{\itemsep}{0ex}}{\end{oldthebibliography}}

\begin{document}
\fontsize{11}{11}\selectfont 
\title{Wave-particle interactions in the outer radiation belts}
\author{\textsl{O.V. Agapitov$^{1,2}$, F.S. Mozer$^{1}$, A. V. Artemyev $^{3}$ , D. Mourenas$^{4}$, V. V. Krasnoselskikh$^{5}$}}
\date{\vspace*{-6ex}}
\maketitle
\begin{center} {\small $^{1}$Space Science Laboratory, the University of California 7 Gauss Way, Berkeley, CA 94720, USA\\
$^{2}$Taras Shevchenko National University of Kyiv, Glushkova ave., 4, 03127, Kyiv, Ukraine\\
$^{3}$University of California Los Angeles, Los Angeles, CA, USA\\
$^{4}$CEA, DAM, DIF, Arpajon, France\\
$^{5}$LPC2E/CNRS-University of Orleans, France\\
{\tt agapitov@ssl.berkeley.edu, agapit@univ.kiev.ua}}\\
\end{center}

\begin{abstract}
Data from the Van Allen Probes have provided the first extensive evidence of non-linear (as opposed to quasi-linear) wave-particle interactions in space with the associated rapid (fraction of  a bounce period) electron acceleration to hundreds of keV by Landau resonance in the parallel electric fields of time domain structures (TDS) and very oblique chorus waves.  The experimental evidence, simulations, and theories of these processes are discussed. {\bf Key words:} the radiation belts, wave-particle interaction, plasma waves and instabilities
\end{abstract}

Understanding the formation, dynamics, and loss of particles in the Earth's radiation belts is a problem that has been studied for decades and that remains incomplete \cite[]{Horne05Nature, Thorne10:Natur, Thorne13:nature}.The Van Allen radiation belt particle populations exist in a dynamic equilibrium between losses (mainly due to particle precipitation to the upper atmosphere) and re-filling due to external injections, transport and acceleration processes \cite[]{Lyons&Thorne73}. The recent results from the Van Allen Probes support the local nature of particle acceleration \cite[]{Reeves13, Mozer14}, and it is now generally agreed that in-situ local acceleration mechanisms are important to the outer radiation belt.

A favored candidate for driving in-situ acceleration (and scattering) is the interaction between whistler waves and electrons \cite[]{Horne05Nature, Thorne10:Natur, Thorne13:nature}. Whistlers in the magnetosphere manifest themselves as electromagnetic perturbations generated near the geomagnetic equator and called chorus (structured, coherent, wave packets \cite[]{Tsurutani&Smith74}) and hiss (diffusive, wide band emissions mostly with randomized phases \cite[]{1973JGR....78.1581T, Meredith_etal2004}). Chorus waves are the most intense natural VLF electromagnetic emissions in the magnetosphere. The maximum chorus wave amplitudes are located in the core of the outer Van Allen radiation belt \cite[]{Agapitov13:jgr}. This fact, as well as the closeness of their frequency to the electron cyclotron frequency, make them one of the key factors that control the dynamics of the outer radiation belt \cite[]{Horne05Nature, Thorne10:Natur, Thorne13:nature, Reeves13}.

Interactions between whistler waves and electrons are analyzed in two approximations:

1. The quasi-linear approach, where particles interact with large numbers of small amplitude waves having random phases and the characteristic time scales are hours and days. The quasi-linear approach to wave-particle interactions was developed in \cite[]{Kennel&Engelmann66, Kennel&Petschek66, Trakhtengerts66}, and applied to the radiation belts in \cite[]{Lyons72} with inclusion of diffusion in particle energy and pitch-angle space \cite[]{Albert07, Glauert&Horne05, Horne_etal2005JGRA, Shprits07, Shprits06, Shprits_etal, Artemyev13:angeo, Agapitov14:jgr:acceleration, Mourenas14} and radial transport of electrons from the tail \cite[]{SchulzLanzerotti1974book, Lanzerotti80}. For the moment, long time prognoses and models of radiation belts dynamics are provided in a frame of this quasi-linear approximation \cite[]{Shprits_etal, Thorne13:nature}. The averaged ($\sim1$ s and more) spectral measurements of magnetic field perturbation with low resolution plasma density measurements (from Cluster, THEMIS, Van Allen Probes missions, see details below) are good enough to provide the experimental basis for such models \cite[]{Shprits08:JASTP_local, Meredith09, Artemyev13:angeo, Mourenas12:JGR, Agapitov14:jgr:AKEBONO}. The problems related to whistler generation and fast anisotropy relaxation of hot electrons population, however, remained out of scope of interests, and the results are mostly obtained from the numerical self-consistent calculations on the basis of Particle-In-Sell (PIC) models. These results indicate large initial anisotropy of electron temperature (greater than 2 for non-thermal population, that has not been observed in averaged data) and its fast relaxation (during $\sim10$ gyroperiods) to values less about 2 \cite[]{Drake15}. The recent results from Cluster mission showed that significant part of whistler waves in the outer radiation belts is oblique \cite[]{Agapitov12:GRL:corrections, Agapitov11:GRL,Agapitov13:jgr, Artemyev15:natcom}. These results were confirmed by numerical simulations based on ray tracing approximation \cite[]{Bortnik_etal2011, Breuillard12:angeo, Chen13, Breuillard15:angeo}. Presence of oblique waves sufficiently affect scattering processes decreasing life time of electrons of near relativistic and relativistic energies \cite[]{Artemyev12:GRL, Mourenas12:JGR, Mourenas12:JGR:acceleration, Artemyev13:angeo, Artemyev12:jgr:distribution, Artemyev15:natcom}. The dependence of chorus wave obliqueness on the geomagnetic activity explains well the geomagnetic storm dynamics of the electron population in the outer radiation belt \cite[]{Artemyev13:grl, Mourenas14, Artemyev15:natcom}. Questions of generation of very oblique waves are considered in \cite[]{Mourenas15}.

2. The non-linear approach where particles can be trapped into the effective potential of large amplitude coherent waves and the characteristic scales of interactions could be fraction of the bounce period or even couple of gyroperiods. Discovery of large electric field amplitudes in whistler waves, above 100 mV/m \cite[]{Cattell_etal2008, Cully_etal2008, Kellogg11, Wilson12, Agapitov14:jgr:acceleration} and numerous observation of large amplitude whistlers by Van Allen Probes, caused renewed interest in non-linear effects in wave-particle interactions. The interaction (electron acceleration) of electrons with coherent whistlers was studied as due to the non-linear cyclotron resonance of electrons moving in the opposite direction of a parallel whistler wave to satisfy the resonance condition \cite[]{Trakhtengerts03, Omura07, Bortnik08} (the concept of such interaction is well established analytically and through the numerical simulation but just several cases from the observations, see for example \cite[]{Mozer14}). The temporal and spatial scales of chorus source and wave coherence are important parameters for feasibility of nonlinear interaction processes, were studied in \cite[]{Santolik05, Agapitov_etal2011jgr, Agapitov_etal2010AnGeo}. The source region scale was found to be around 600 km at $L\sim4.5$ from Cluster measurements and about 3000 km from measurements of five THEMIS spacecraft \cite[]{Agapitov_etal2010AnGeo}. Numerous observations of similar chorus wave packets  aboard Van Allen Probes spacecraft at separations greater than 500 km indicate that the chorus source spatial scales varies from 500 to 1000 km in the outer radiation belt \cite[]{Agapitov_etal2011jgr}. Thus, recent results show that the scale of the source region for chorus waves is much greater than the wavelength ($\sim10-20$ wavelengths) and that this scale may depend strongly on $L$-shell. The temporal whistler wave coherence is discussed in the recent paper by \cite[]{Gao_etal2014}, where the high level of time coherence was obtained for chorus waves (rising and falling tones) but a much lower coherence level was obtained for hiss. The detailed study of nonlinear trapping and acceleration as well as of breaking the resonance conditions for cyclotron and Landau resonances is presented in \cite[]{Artemyev15:pop:stability}.

It was shown that the majority of the observed large amplitude waves were very oblique \cite[]{Cattell_etal2008, Agapitov14:jgr:acceleration} and thus had large electric field, also, aligned to the background magnetic field (see figure 4 in \cite[]{Agapitov14:jgr:acceleration}). Large electric field component parallel to the background magnetic field allows these waves to interact with electrons through the Landau resonance that reduces the energy threshold for non-linear trapping to $\sim1$ keV \cite[]{Artemyev12:pop:nondiffusion, Agapitov14:jgr:acceleration}. This interaction provides acceleration along the background magnetic field line, which can cause precipitation of $10-100$ keV electrons \cite[]{Artemyev12:pop:nondiffusion} that presumably becomes the source for x-ray micro-bursts. Moreover, such interaction provides the significant feedback for wave fields and tends to the nonlinear transformation of waveforms that was demonstrated on the basis of electric field measurements on board the Van Allen Probes \cite[]{Mozer14} making the dramatic changes for distributing function on time scales about couple of gyroperiods due to trapping. The Landau trapping responsible for such interactions needs high resolution particles data for confirmation (all results were obtained from the numerical PIC simulation \cite[]{Drake15}) and these processes are the subject for future projects. The Landau resonance trapping can provide the feedback response from trapped particle to the wave fields and Van Allen Probes project presents the high quality data for electromagnetic fields and particles to study in details these nonlinear effects.

Data from the Van Allen Probes have provided the first extensive evidence of nonlinear (as opposed to quasi-linear) wave-particle interactions in space with the associated rapid (about a bounce period) electron acceleration to tens (and hundreds in some cases) of keV by the Landau resonance in the parallel electric fields of time domain structures (TDS) \cite[]{Artemyev14:grl:thermal, Vasko2015JGR, Mozer14} and very oblique chorus waves \cite[]{Agapitov15:grl2}. The Van Allen Probes A and B were launched on August 30, 2012, aiming at studying the radiation belts with unprecedented detail \cite[]{Mauk13}. The data is provided by the following instruments on board the Van Allen Probes: the Electric Field and Waves (EFW) instrument \citep[]{Wygant13} for electric field waveforms; the Electric and Magnetic Field Instrument Suite and Integrated Science (EMFISIS) \citep[]{Kletzing13} for electric and magnetic spectral and polarization information, the DC magnetic field data and electron cyclotron frequency $f_{ce}$, the electron plasma frequency value $f_{pe}$ (deduced from the upper hybrid frequency); the Energetic Particle, Composition, and Thermal Plasma (ECT) instrument \citep{Spence13} data of the Helium Oxygen Proton Electron (HOPE) detector for electron fluxes below 50 keV \citep{Funsten13}, and the  Magnetic Electron Ion Spectrometer (MagEIS) 30 keV to 4 MeV measurements \citep{Blake13}. The waveform data was transmitted at 16,384 samples/s (35,000 from EMFISIS) for 5 s continuous intervals in the burst mode (triggered automatically by large amplitude signals), as well as continuously every 6 s under the form of full spectral matrices for 65 frequencies (logarithmically spaced from 3 Hz to 16 kHz for EMFISIS).

\begin{figure}
\includegraphics[width=40pc]{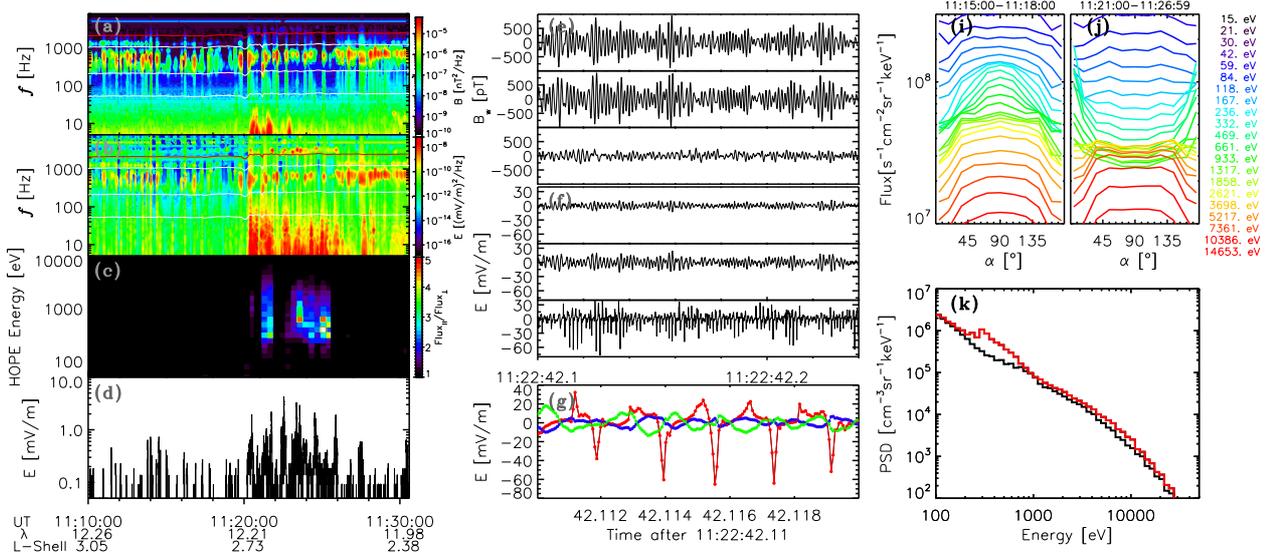}
\caption{Observations of nonlinear whistler waves on Van Allen Probe B spacecraft on May 1, 2013 : (a,b) magnetic and electric field spectrograms; (c) the ratio
of parallel and perpendicular electron fluxes at different energies (see text for details); (d) the average amplitude of the electric field fluctuations (averaged for <300 Hz fluctuations);
(e,f) waveforms of magnetic and electric fields captured during the period of enhanced electric field fluctuations; (g) one of electric field spikes presented in panel (f); (h,i) electron pitch angle distributions observed during weak and intense electric field fluctuations; (j) field-aligned phase space densities corresponding (h) and (i). } \label{fig1}
\end{figure}

Figure \ref{fig1} provides an example of these nonlinear processes.  Panels (a) and (b) give the spectra of the magnetic and electric field during a 20 minute interval.  The red horizontal line in each of these panels gives the electron gyrofrequency and the white line below it is half this frequency.  Thus, the waves seen at $300-1000$ Hz are lower band chorus.  Panel (c) gives the flux ratio, defined as the average of the electron number flux at the two lowest and two highest pitch angles divided by the 90 degree flux.   Flux ratios of $2-4$ (field-aligned flux two or four times greater than the 90 degree flux) are observed for $\sim1$ keV electrons between about 1120 and 1126 UT, while chorus waves were measured by the wave detectors.  Panels (e), (f), and (g) give the three components of the magnetic field in field-aligned coordinates while panels (h), (i), and (j) give the same information for the electric field.  The parallel components of the magnetic and electric fields are in panels (g) and (j), respectively.  In panel (j), it is seen that the parallel electric field is distorted to produce periodic pulses of parallel electric field that are viewed in the expanded plot of panel (k).  Panels (l) and (m) give the electron pitch angle distributions before and during the time of the non-linearity.  They show that the 100 eV to 5 keV electrons (the green and yellow curves) became field-aligned as a result of this non-linear interaction.  Thus, this figure provides an example of the non-linear interaction between a chorus wave and the electron distribution that results in a distorted parallel electric field (panel j) and field-aligned acceleration of keV electrons by this distorted field (panel m) .

Formation of TDS from the background fields and plasmas is discussed in \cite[]{Drake15, Vasko15:grl,agapitov2015grl:param}. TDS structures are observed on the night side at $L$ values of $\sim5$ during perturbed conditions in the magnetosphere. They coincide with magnetic energy injections and field-aligned currents and, sometimes, with injected particles, although injected particles may not frequently penetrate to these $L$-shells. A statistical analysis will be performed to determine if a correlation between field-aligned currents and TDS exists in the Van Allen Probe measurements. During the event reported by \cite[]{Mozer14} and by \cite[]{agapitov2015grl:param}, whistler waves and TDS structures were simultaneously present. Moreover, more detailed analyses have shown that these two types of structures can be phase correlated. This suggests that some TDS structures can be directly generated by wave-wave (as it was shown by \cite[]{agapitov2015grl:param}) and/or wave-particle interactions with whistlers and preliminary simulations support such a hypothesis. Nonlinear whistler wave-wave and wave-particle interactions will be studied to determine the role of these waves in the generation and evolution of TDS and modification of the particle distributions.
TDS generation may be related to the large amplitude whistler waves that are frequently observed in the outer radiation belt \cite[]{Agapitov14:jgr:acceleration}. Preliminary studies have shown that such whistlers are mainly very oblique with a significant parallel electric field component. This field can trap and accelerate electrons (see \cite[]{Artemyev13:pop, Artemyev12:pop:nondiffusion} for details) and it can produce non-linear effects on the wave \cite[]{Kellogg11, Mozer14}. Such non-linear steepening of whistlers can further manifest itself in growth of upper wave harmonics and strong localization in space with a consequent increase of the parallel electric field perturbation. The statistics of large amplitude whistler waves with the statistics of different types of TDS will be generated. PIC simulations \cite[]{Drake15} suggested that the parallel electric fields of off-angle whistlers trigger the formation of intense bipolar electric field structures and double layers.

\begin{figure}
\includegraphics[width=40pc]{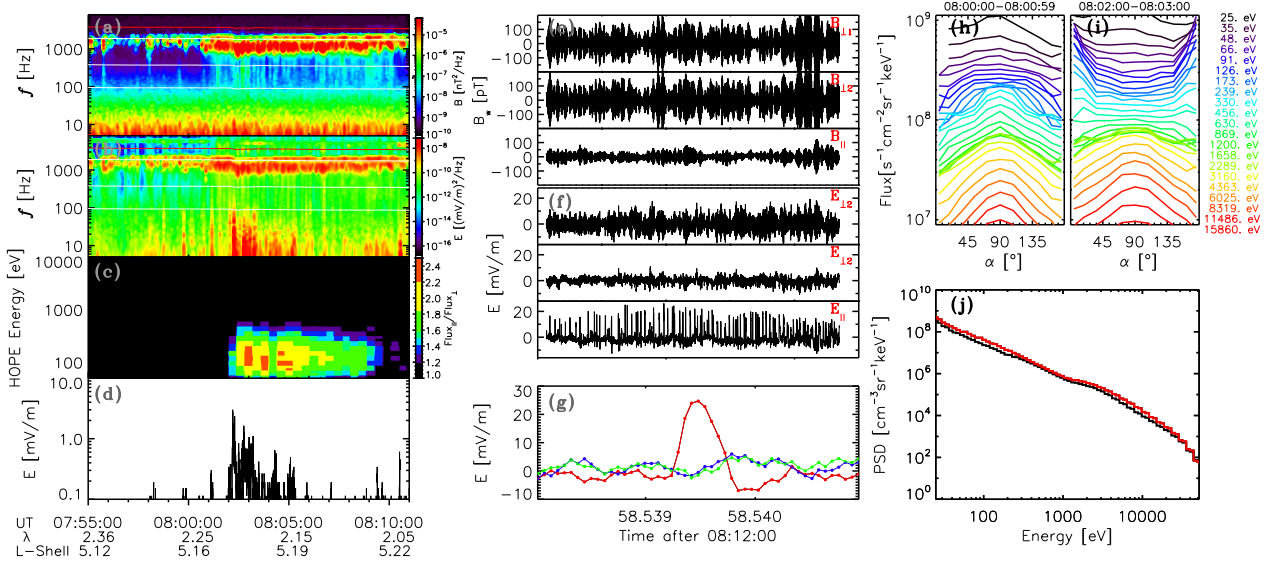}
\caption{ Observations of field-aligned electron PADs conjugated with observation of
TDSs on November 1, 2012. The data format is the same as for Fig. 1.} \label{fig2}
\end{figure}

An example of TDS and parallel electron acceleration is given in figure 2 from Van Allen Probe B measurements on $2012-11-01$ (panels are similar to those in figure 1). Field-aligned electrons with energies as high as hundreds of eV are seen in the flux ratio plot of panel (c), while TDS, identified by the broadband electrostatic noise, are observed in panel (b). The free energy source for the TDS and the field-aligned electrons might be the field-aligned current and injections -- such current systems are often observed together with TDS \cite[]{Malaspina15, Malaspina14}. Examples of pitch angle distributions before and during a TDS event are presented in panels (h) and (i), showing the field-alignment of 500-5000 eV electrons. It is noted, that the injected electrons did not have field-aligned distributions, which is evidence that such distributions were made by the TDS. The parallel electric field of the TDS is shown in panel (f) (second plot from the bottom of the central figure) and the parallel magnetic field is given in panel (e) (third panel from the top of the central figure). The fine structure of the single TDS is shown in panel (g). These plots show that the observed TDS are the electrostatic double layers with amplitudes of $\sim30$ mV/m in the parallel electric field component. TDS velocity is estimated to be about 3000 km/s. Statistics of TDS distribution in the outer radiation belts and association with other phenomena of geomagnetic activity \cite[]{Malaspina15, Malaspina14}

The role of TDS in the acceleration of thermal electrons is discussed in \cite[]{Artemyev14:grl:thermal, Vasko15:jgr:DL}. The interaction of TDS with thermal electrons (\cite[]{Artemyev14:grl:thermal} and \cite[]{Mozer14}) causes parallel electron acceleration to keV energies. This is important for developing the seed population required for the further relativistic acceleration and TDS may thus be important for the dynamics of the radiation belts. To test whether TDS provide this seed population, a systematic approach is planned, based on statistics. The objective is to compare examples of the effect on thermal electrons of whistlers with TDS, whistlers without TDS and TDS without whistlers. Such data will depend on time domain waveforms collected and transmitted to ground. Time domain data is required because power spectra alone do not provide sufficient information on the obliqueness of the whistlers or the type of TDS. For example, a low frequency signal in the power spectrum could be due to TDS or to lower hybrid waves or to neither, so time domain data is required.

High-time resolution measurements, such as those presented in Figure 1 and 2, have thus far been analyzed during $\sim5$ second bursts at a duty cycle of about 1\% of the time from both the Van Allen Probes and THEMIS. While this data provides a peek into the relevant processes, it is planned to obtain a microscopic view of the physics by analyzing hours of continuous high time resolution plasma and field data from the Van Allen Probes during night-time apogees in 2014 through 2015 and from THEMIS during dedicated high data rate bursts of tens of minutes each in 2015 and 2016.  Many of these burst collections will be done in conjunction with the Canadian ground based array of auroral images in order to study the physics that leads to particle precipitation and auroras. The apogee of the Van Allen probes passed through dawn and to 2200 MLT by late summer of 2015. During this time interval, the apogee passed through the in-situ local acceleration region near midnight, which is the locale of the largest TDS concentration. The electric field instrument operated in the new mode in which continuous measurements at a sample rate of 16384 samples/second for $3-5$ hour intervals were made to supplement the 5 second bursts.  Similarly, during $2014-2015$, tens of examples of seven minute bursts of plasma and field data were collected on THEMIS at times of conjunctions between the satellite and one or more Canadian ground stations. Accordingly, a systematic approach will be performed in which the high-time resolution observations gathered by the Van Allen Probes and THEMIS will be subjected to theoretical interpretation, modeling, and computer simulation.

The model of TDS interactions with thermal electrons has been developed in \cite[]{Artemyev14:grl:thermal, Vasko15:jgr:DL} and the results reproduce the main features of particle dynamics obtained from the Van Allen Probe measurements (figures 1 and 2). The multiple interactions with TDS affect mostly particles below 1 keV but trapping effects can accelerate electrons to $10-20$ keV.

The influence of TDS in relativistic electron acceleration and scattering is studied in \cite[]{Mozer14, Mozer15}. Low energy electrons can be accelerated to relativistic energies by the two-step process of keV acceleration by a small number of resonant interactions with TDS \cite[]{Artemyev14:grl:thermal} followed by whistler cyclotron resonance acceleration, as was shown experimentally and by a test particle simulation in \cite[]{Mozer14}. A multi-case study of this two-step process will be carried out through a statistical analysis of electron distribution function, TDS, and chorus wave parameters in TDS and whistler events in order to determine the key factors controlling these dynamics. An important issue is the role of upper band whistlers because the cyclotron resonance condition is satisfied by lower energy electrons as the wave frequency approaches the electron cyclotron frequency. Upper band waves (with frequencies from 0.55 to 0.8 of the local electron gyrofrequency) are generally assumed to be less intense than lower band chorus whistlers (with frequencies from 0.1 to $\sim0.45$ of the local electron gyrofrequency) but statistical studies of upper band chorus have been limited by instrumental frequency responses ($\sim4$ kHz on Cluster and THEMIS). One of the proposed tasks will be the detailed study of upper band whistlers to consider their occurrence frequency and role in particle acceleration (on the Van Allen Probes the high rate time domain data has a frequency response to 8 kHz.).
Not considered in the two-step process of relativistic electron acceleration is the possibility of additional acceleration by the Landau resonance with oblique whistlers \cite[]{Artemyev12:pop:nondiffusion}. This mechanism requires a low threshold in electron energy ($\sim2-10$ keV) to satisfy the trapping conditions and it can function efficiently to accelerate (up to 100 keV) and scatter electrons after their pre-acceleration by TDS.

Observations by the Van Allen Probes have revealed nonlinear features of the ELF/VLF waves in the outer radiation belt that can be important for particle diffusion and acceleration. The short review of the recent publications dedicated to nonlinear wave-particles interactions is presented here. To conclude it has been shown that during local night time at $L\sim5$, multiple electric field bursts of short duration (several hundred micro-seconds) and large amplitudes (up to 300 mV/m), called time domain structures (TDS) \cite[]{Mozer13, Mozer15}, are observed \cite[]{Artemyev14:grl:thermal, Mozer14} to correlate with energetic particles having magnetic-field-aligned pitch angle distributions. Such TDS can provide the first step \cite[]{Artemyev14:grl:thermal, Vasko2015JGR} in a two-step process that accelerates thermal electrons to relativistic energies \cite[]{Mozer14}. The second step in the process is perpendicular acceleration by cyclotron resonance with whistlers. Thermal particles cannot be accelerated by such chorus waves because they do not satisfy the resonance condition (see Section 1.2 for details). Injected particles may not penetrate to the required depth for them to become a significant source of seed particles for the cyclotron \cite[]{Mozer14} and/or the Landau resonance \cite[]{Artemyev12:pop:nondiffusion}. Because TDS acceleration leads to an increase of the electron parallel velocity (see figure 3 in \cite[]{Artemyev14:grl:thermal}) and because the whistler resonance threshold depends on this parallel velocity, the TDS interaction makes it possible for thermal electrons to be accelerated to relativistic energies through the chain of trapping \cite[]{Mozer14}. The simultaneous observations of whistler waves and TDS along with relativistic accelerated electrons provide clear evidence that this two-step process is active. Additional evidence is that TDS are observed to accelerate electrons to keV energies in the absence of whistlers and that whistlers alone do not accelerate electrons. These facts imply that two steps are crucial for acceleration of electrons from thermal to relativistic energies.
\section*{\sc acknowledgement}
\indent \indent The work by O.A. was performed under JHU/APL Contract No. 922613 (RBSP-EFW).

\bibliographystyle{natbib}

\begin{thebibliography}{72}
\expandafter\ifx\csname natexlab\endcsname\relax\def\natexlab#1{#1}\fi
\expandafter\ifx\csname url\endcsname\relax
  \def\url#1{\texttt{#1}}\fi
\expandafter\ifx\csname urlprefix\endcsname\relax\def\urlprefix{URL }\fi

\bibitem[{{Agapitov} \emph{et~al.}(2010){Agapitov}, {Krasnoselskikh},
  {Zaliznyak}, {Angelopoulos}, {Le Contel}, and
  {Rolland}}]{Agapitov_etal2010AnGeo}
{Agapitov}, O., {Krasnoselskikh}, V., {Zaliznyak}, Y., {Angelopoulos}, V., {Le
  Contel}, O., and {Rolland}, G. (2010).
\newblock {Chorus source region localization in the Earth's outer magnetosphere
  using THEMIS measurements}.
\newblock \emph{Annales Geophysicae}, \textbf{28}, 1377--1386.

\bibitem[{{Agapitov} \emph{et~al.}(2011{\natexlab{a}}){Agapitov},
  {Krasnoselskikh}, {Dudok de Wit}, {Khotyaintsev}, {Pickett},
  {Santol{\'{\i}}k}, and {Rolland}}]{Agapitov_etal2011jgr}
{Agapitov}, O., {Krasnoselskikh}, V., {Dudok de Wit}, T., {Khotyaintsev}, Y.,
  {Pickett}, J.~S., {Santol{\'{\i}}k}, O., and {Rolland}, G.
  (2011{\natexlab{a}}).
\newblock {Multispacecraft observations of chorus emissions as a tool for the
  plasma density fluctuations' remote sensing}.
\newblock \emph{Journal of Geophysical Research (Space Physics)}, \textbf{116},
  9222.

\bibitem[{{Agapitov} \emph{et~al.}(2011{\natexlab{b}}){Agapitov},
  {Krasnoselskikh}, {Khotyaintsev}, and {Rolland}}]{Agapitov11:GRL}
{Agapitov}, O., {Krasnoselskikh}, V., {Khotyaintsev}, Y.~V., and {Rolland}, G.
  (2011{\natexlab{b}}).
\newblock {A statistical study of the propagation characteristics of whistler
  waves observed by Cluster}.
\newblock \emph{\grl}, \textbf{382}, 20\,103.

\bibitem[{{Agapitov} \emph{et~al.}(2012){Agapitov}, {Krasnoselskikh},
  {Khotyaintsev}, and {Rolland}}]{Agapitov12:GRL:corrections}
{Agapitov}, O., {Krasnoselskikh}, V., {Khotyaintsev}, Y.~V., and {Rolland}, G.
  (2012).
\newblock {Correction to "A statistical study of the propagation
  characteristics of whistler waves observed by Cluster"}.
\newblock \emph{\grl}, \textbf{39}, 24\,102.

\bibitem[{{Agapitov} \emph{et~al.}(2013){Agapitov}, {Artemyev},
  {Krasnoselskikh}, {Khotyaintsev}, {Mourenas}, {Breuillard}, {Balikhin}, and
  {Rolland}}]{Agapitov13:jgr}
{Agapitov}, O., {Artemyev}, A., {Krasnoselskikh}, V., {Khotyaintsev}, Y.~V.,
  {Mourenas}, D., {Breuillard}, H., {Balikhin}, M., and {Rolland}, G. (2013).
\newblock {Statistics of whistler mode waves in the outer radiation belt:
  Cluster STAFF-SA measurements}.
\newblock \emph{\jgr}, \textbf{118}, 3407--3420.

\bibitem[{{Agapitov} \emph{et~al.}(2014{\natexlab{a}}){Agapitov}, {Artemyev},
  {Mourenas}, {Krasnoselskikh}, {Bonnell}, {Le Contel}, {Cully}, and
  {Angelopoulos}}]{Agapitov14:jgr:acceleration}
{Agapitov}, O., {Artemyev}, A., {Mourenas}, D., {Krasnoselskikh}, V.,
  {Bonnell}, J., {Le Contel}, O., {Cully}, C.~M., and {Angelopoulos}, V.
  (2014{\natexlab{a}}).
\newblock {The quasi-electrostatic mode of chorus waves and electron nonlinear
  acceleration}.
\newblock \emph{\jgr}, \textbf{119}, 1606–1626.

\bibitem[{{Agapitov} \emph{et~al.}(2015{\natexlab{a}}){Agapitov}, {Artemyev},
  {Mourenas}, {Mozer}, and {Krasnoselskikh}}]{Agapitov15:grl2}
{Agapitov}, O., {Artemyev}, A., {Mourenas}, D., {Mozer}, F., and
  {Krasnoselskikh}, V. (2015{\natexlab{a}}).
\newblock {Nonlinear local parallel acceleration of electrons through Landau
  trapping by oblique whistler-mode waves in the outer radiation belt}.
\newblock \emph{\grl}, page submitted.

\bibitem[{{Agapitov} \emph{et~al.}(2014{\natexlab{b}}){Agapitov}, {Artemyev},
  {Mourenas}, {Kasahara}, and {Krasnoselskikh}}]{Agapitov14:jgr:AKEBONO}
{Agapitov}, O.~V., {Artemyev}, A.~V., {Mourenas}, D., {Kasahara}, Y., and
  {Krasnoselskikh}, V. (2014{\natexlab{b}}).
\newblock {Inner belt and slot region electron lifetimes and energization rates
  based on AKEBONO statistics of whistler waves}.
\newblock \emph{\jgr}, \textbf{119}, 2876--2893.

\bibitem[{{Agapitov} \emph{et~al.}(2015{\natexlab{b}}){Agapitov},
  {Krasnoselskikh}, {Mozer}, {Artemyev}, and
  {Volokitin}}]{agapitov2015grl:param}
{Agapitov}, O.~V., {Krasnoselskikh}, V., {Mozer}, F.~S., {Artemyev}, A.~V., and
  {Volokitin}, A.~S. (2015{\natexlab{b}}).
\newblock {Generation of nonlinear electric field bursts in the outer radiation
  belt through the parametric decay of whistler waves}.
\newblock \emph{\grl}, \textbf{42}, 3715--3722.

\bibitem[{{Albert}(2007)}]{Albert07}
{Albert}, J.~M. (2007).
\newblock {Simple approximations of quasi-linear diffusion coefficients}.
\newblock \emph{J. Geophys. Res.}, \textbf{112}, 12\,202.

\bibitem[{{Artemyev} \emph{et~al.}(2012{\natexlab{a}}){Artemyev}, {Agapitov},
  {Breuillard}, {Krasnoselskikh}, and {Rolland}}]{Artemyev12:GRL}
{Artemyev}, A., {Agapitov}, O., {Breuillard}, H., {Krasnoselskikh}, V., and
  {Rolland}, G. (2012{\natexlab{a}}).
\newblock {Electron pitch-angle diffusion in radiation belts: The effects of
  whistler wave oblique propagation}.
\newblock \emph{\grl}, \textbf{39}, 8105.

\bibitem[{{Artemyev} \emph{et~al.}(2012{\natexlab{b}}){Artemyev}, {Agapitov},
  {Krasnoselskikh}, {Breuillard}, and {Rolland}}]{Artemyev12:jgr:distribution}
{Artemyev}, A., {Agapitov}, O., {Krasnoselskikh}, V., {Breuillard}, H., and
  {Rolland}, G. (2012{\natexlab{b}}).
\newblock {Statistical model of electron pitch-angle diffusion in the outer
  radiation belt.}
\newblock \emph{J. Geophys. Res.}, \textbf{117}, A08\,219.

\bibitem[{{Artemyev} \emph{et~al.}(2012{\natexlab{c}}){Artemyev},
  {Krasnoselskikh}, {Agapitov}, {Mourenas}, and
  {Rolland}}]{Artemyev12:pop:nondiffusion}
{Artemyev}, A., {Krasnoselskikh}, V., {Agapitov}, O., {Mourenas}, D., and
  {Rolland}, G. (2012{\natexlab{c}}).
\newblock {Non-diffusive resonant acceleration of electrons in the radiation
  belts.}
\newblock \emph{Physics of Plasmas}, \textbf{19}, 122\,901.

\bibitem[{{Artemyev} \emph{et~al.}(2013{\natexlab{a}}){Artemyev}, {Agapitov},
  {Mourenas}, {Krasnoselskikh}, and {Zelenyi}}]{Artemyev13:grl}
{Artemyev}, A.~V., {Agapitov}, O.~V., {Mourenas}, D., {Krasnoselskikh}, V., and
  {Zelenyi}, L.~M. (2013{\natexlab{a}}).
\newblock {Storm-induced energization of radiation belt electrons: Effect of
  wave obliquity}.
\newblock \emph{\grl}, \textbf{40}, 4138--4143.

\bibitem[{{Artemyev} \emph{et~al.}(2013{\natexlab{b}}){Artemyev}, {Mourenas},
  {Agapitov}, and {Krasnoselskikh}}]{Artemyev13:angeo}
{Artemyev}, A.~V., {Mourenas}, D., {Agapitov}, O.~V., and {Krasnoselskikh},
  V.~V. (2013{\natexlab{b}}).
\newblock {Parametric validations of analytical lifetime estimates for
  radiation belt electron diffusion by whistler waves}.
\newblock \emph{Annales Geophysicae}, \textbf{31}, 599--624.

\bibitem[{{Artemyev} \emph{et~al.}(2013{\natexlab{c}}){Artemyev}, {Vasiliev},
  {Mourenas}, {Agapitov}, and {Krasnoselskikh}}]{Artemyev13:pop}
{Artemyev}, A.~V., {Vasiliev}, A.~A., {Mourenas}, D., {Agapitov}, O., and
  {Krasnoselskikh}, V. (2013{\natexlab{c}}).
\newblock {Nonlinear electron acceleration by oblique whistler waves: Landau
  resonance vs. cyclotron resonance.}
\newblock \emph{Physics of Plasmas}, \textbf{20}, 122\,901.

\bibitem[{{Artemyev} \emph{et~al.}(2014){Artemyev}, {Agapitov}, {Mozer}, and
  {Krasnoselskikh}}]{Artemyev14:grl:thermal}
{Artemyev}, A.~V., {Agapitov}, O., {Mozer}, F., and {Krasnoselskikh}, V.
  (2014).
\newblock {Thermal electron acceleration by localized bursts of electric field
  in the radiation belts}.
\newblock \emph{\grl}, \textbf{41}, 5734–5739.

\bibitem[{{Artemyev} \emph{et~al.}(2015{\natexlab{a}}){Artemyev}, {Agapitov},
  {Mourenas}, {Krasnoselskikh}, and {Mozer}}]{Artemyev15:natcom}
{Artemyev}, A.~V., {Agapitov}, O.~V., {Mourenas}, D., {Krasnoselskikh}, V.~V.,
  and {Mozer}, F.~S. (2015{\natexlab{a}}).
\newblock {Wave energy budget analysis in the Earth's radiation belts uncovers
  a missing energy}.
\newblock \emph{Nature Communications}, \textbf{6}, 8143.

\bibitem[{{Artemyev} \emph{et~al.}(2015{\natexlab{b}}){Artemyev}, {Mourenas},
  {Agapitov}, {Vainchtein}, {Mozer}, and
  {Krasnoselskikh}}]{Artemyev15:pop:stability}
{Artemyev}, A.~V., {Mourenas}, D., {Agapitov}, O.~V., {Vainchtein}, D.~L.,
  {Mozer}, F.~S., and {Krasnoselskikh}, V.~V. (2015{\natexlab{b}}).
\newblock {Stability of relativistic electron trapping by strong whistler or
  electromagnetic ion cyclotron waves}.
\newblock \emph{Physics of Plasmas}, \textbf{22}, 082\,901.

\bibitem[{{Blake} \emph{et~al.}(2013){Blake}, {Carranza}, {Claudepierre},
  {Clemmons}, {Crain}, {Dotan}, {Fennell}, {Fuentes}, {Galvan}, {George},
  {Henderson}, {Lalic}, {Lin}, {Looper}, {Mabry}, {Mazur}, {McCarthy},
  {Nguyen}, {O'Brien}, {Perez}, {Redding}, {Roeder}, {Salvaggio}, {Sorensen},
  {Spence}, {Yi}, and {Zakrzewski}}]{Blake13}
{Blake}, J.~B., {Carranza}, P.~A., {Claudepierre}, S.~G., {Clemmons}, J.~H.,
  {Crain}, W.~R., {Dotan}, Y., {Fennell}, J.~F., {Fuentes}, F.~H., {Galvan},
  R.~M., {George}, J.~S., {Henderson}, M.~G., {Lalic}, M., {Lin}, A.~Y.,
  {Looper}, M.~D., {Mabry}, D.~J., {Mazur}, J.~E., {McCarthy}, B., {Nguyen},
  C.~Q., {O'Brien}, T.~P., {Perez}, M.~A., {Redding}, M.~T., {Roeder}, J.~L.,
  {Salvaggio}, D.~J., {Sorensen}, G.~A., {Spence}, H.~E., {Yi}, S., and
  {Zakrzewski}, M.~P. (2013).
\newblock {The Magnetic Electron Ion Spectrometer (MagEIS) Instruments Aboard
  the Radiation Belt Storm Probes (RBSP) Spacecraft}.
\newblock \emph{\ssr}, \textbf{179}, 383--421.

\bibitem[{{Bortnik} \emph{et~al.}(2008){Bortnik}, {Thorne}, and
  {Inan}}]{Bortnik08}
{Bortnik}, J., {Thorne}, R.~M., and {Inan}, U.~S. (2008).
\newblock {Nonlinear interaction of energetic electrons with large amplitude
  chorus}.
\newblock \emph{\grl}, \textbf{35}, 21\,102.

\bibitem[{{Bortnik} \emph{et~al.}(2011){Bortnik}, {Chen}, {Li}, {Thorne},
  {Meredith}, and {Horne}}]{Bortnik_etal2011}
{Bortnik}, J., {Chen}, L., {Li}, W., {Thorne}, R.~M., {Meredith}, N.~P., and
  {Horne}, R.~B. (2011).
\newblock {Modeling the wave power distribution and characteristics of
  plasmaspheric hiss}.
\newblock \emph{Journal of Geophysical Research (Space Physics)}, \textbf{116},
  12\,209.

\bibitem[{{Breuillard} \emph{et~al.}(2012){Breuillard}, {Zaliznyak},
  {Krasnoselskikh}, {Agapitov}, {Artemyev}, and {Rolland}}]{Breuillard12:angeo}
{Breuillard}, H., {Zaliznyak}, Y., {Krasnoselskikh}, V., {Agapitov}, O.,
  {Artemyev}, A., and {Rolland}, G. (2012).
\newblock {Chorus wave-normal statistics in the Earth’s radiation belts from
  ray tracing technique}.
\newblock \emph{Ann. Geophys.}, \textbf{30}, 1223–1233.

\bibitem[{Breuillard \emph{et~al.}(2015)Breuillard, Agapitov, Artemyev,
  Kronberg, Haaland, Daly, Krasnoselskikh, Boscher, Bourdarie, Zaliznyak, and
  Rolland}]{Breuillard15:angeo}
Breuillard, H., Agapitov, O., Artemyev, A., Kronberg, E.~A., Haaland, S.~E.,
  Daly, P.~W., Krasnoselskikh, V.~V., Boscher, D., Bourdarie, S., Zaliznyak,
  Y., and Rolland, G. (2015).
\newblock Field-aligned chorus wave spectral power in earth's outer radiation
  belt.
\newblock \emph{Annales Geophysicae}, \textbf{33}(5), 583--597.
\newblock \urlprefix\url{http://www.ann-geophys.net/33/583/2015/}.

\bibitem[{{Cattell} \emph{et~al.}(2008){Cattell}, {Wygant}, {Goetz}, {Kersten},
  {Kellogg}, {von Rosenvinge}, {Bale}, {Roth}, {Temerin}, {Hudson}, {Mewaldt},
  {Wiedenbeck}, {Maksimovic}, {Ergun}, {Acuna}, and
  {Russell}}]{Cattell_etal2008}
{Cattell}, C., {Wygant}, J.~R., {Goetz}, K., {Kersten}, K., {Kellogg}, P.~J.,
  {von Rosenvinge}, T., {Bale}, S.~D., {Roth}, I., {Temerin}, M., {Hudson},
  M.~K., {Mewaldt}, R.~A., {Wiedenbeck}, M., {Maksimovic}, M., {Ergun}, R.,
  {Acuna}, M., and {Russell}, C.~T. (2008).
\newblock {Discovery of very large amplitude whistler-mode waves in Earth's
  radiation belts}.
\newblock \emph{\grl}, \textbf{35}, 1105.

\bibitem[{{Chen} \emph{et~al.}(2013){Chen}, {Thorne}, {Li}, and
  {Bortnik}}]{Chen13}
{Chen}, L., {Thorne}, R.~M., {Li}, W., and {Bortnik}, J. (2013).
\newblock {Modeling the wave normal distribution of chorus waves}.
\newblock \emph{\jgr}, \textbf{118}, 1074--1088.

\bibitem[{{Cully} \emph{et~al.}(2008){Cully}, {Bonnell}, and
  {Ergun}}]{Cully_etal2008}
{Cully}, C.~M., {Bonnell}, J.~W., and {Ergun}, R.~E. (2008).
\newblock {THEMIS observations of long-lived regions of large-amplitude
  whistler waves in the inner magnetosphere}.
\newblock \emph{\grl}, \textbf{35}, 17.

\bibitem[{{Drake} \emph{et~al.}(2015){Drake}, {Agapitov}, and
  {Mozer}}]{Drake15}
{Drake}, J.~F., {Agapitov}, O.~V., and {Mozer}, F.~S. (2015).
\newblock {The development of a bursty precipitation front with intense
  localized parallel electric fields driven by whistler waves}.
\newblock \emph{\grl}, \textbf{42}, 2563--2570.

\bibitem[{{Funsten} \emph{et~al.}(2013){Funsten}, {Skoug}, {Guthrie},
  {MacDonald}, {Baldonado}, {Harper}, {Henderson}, {Kihara}, {Lake}, {Larsen},
  {Puckett}, {Vigil}, {Friedel}, {Henderson}, {Niehof}, {Reeves}, {Thomsen},
  {Hanley}, {George}, {Jahn}, {Cortinas}, {De Los Santos}, {Dunn}, {Edlund},
  {Ferris}, {Freeman}, {Maple}, {Nunez}, {Taylor}, {Toczynski}, {Urdiales},
  {Spence}, {Cravens}, {Suther}, and {Chen}}]{Funsten13}
{Funsten}, H.~O., {Skoug}, R.~M., {Guthrie}, A.~A., {MacDonald}, E.~A.,
  {Baldonado}, J.~R., {Harper}, R.~W., {Henderson}, K.~C., {Kihara}, K.~H.,
  {Lake}, J.~E., {Larsen}, B.~A., {Puckett}, A.~D., {Vigil}, V.~J., {Friedel},
  R.~H., {Henderson}, M.~G., {Niehof}, J.~T., {Reeves}, G.~D., {Thomsen},
  M.~F., {Hanley}, J.~J., {George}, D.~E., {Jahn}, J.-M., {Cortinas}, S., {De
  Los Santos}, A., {Dunn}, G., {Edlund}, E., {Ferris}, M., {Freeman}, M.,
  {Maple}, M., {Nunez}, C., {Taylor}, T., {Toczynski}, W., {Urdiales}, C.,
  {Spence}, H.~E., {Cravens}, J.~A., {Suther}, L.~L., and {Chen}, J. (2013).
\newblock {Helium, Oxygen, Proton, and Electron (HOPE) Mass Spectrometer for
  the Radiation Belt Storm Probes Mission}.
\newblock \emph{\ssr}, \textbf{179}, 423--484.

\bibitem[{{Gao} \emph{et~al.}(2014){Gao}, {Li}, {Thorne}, {Bortnik},
  {Angelopoulos}, {Lu}, {Tao}, and {Wang}}]{Gao_etal2014}
{Gao}, X., {Li}, W., {Thorne}, R.~M., {Bortnik}, J., {Angelopoulos}, V., {Lu},
  Q., {Tao}, X., and {Wang}, S. (2014).
\newblock {Statistical results describing the bandwidth and coherence
  coefficient of whistler mode waves using THEMIS waveform data}.
\newblock \emph{Journal of Geophysical Research (Space Physics)}, \textbf{119},
  8992--9003.

\bibitem[{{Glauert} and {Horne}(2005)}]{Glauert&Horne05}
{Glauert}, S.~A. and {Horne}, R.~B. (2005).
\newblock {Calculation of pitch angle and energy diffusion coefficients with
  the PADIE code}.
\newblock \emph{J. Geophys. Res.}, \textbf{110}, 4206.

\bibitem[{{Horne} \emph{et~al.}(2005{\natexlab{a}}){Horne}, {Thorne},
  {Glauert}, {Albert}, {Meredith}, and {Anderson}}]{Horne_etal2005JGRA}
{Horne}, R.~B., {Thorne}, R.~M., {Glauert}, S.~A., {Albert}, J.~M., {Meredith},
  N.~P., and {Anderson}, R.~R. (2005{\natexlab{a}}).
\newblock {Timescale for radiation belt electron acceleration by whistler mode
  chorus waves}.
\newblock \emph{Journal of Geophysical Research (Space Physics)}, \textbf{110},
  3225.

\bibitem[{{Horne} \emph{et~al.}(2005{\natexlab{b}}){Horne}, {Thorne},
  {Shprits}, {Meredith}, {Glauert}, {Smith}, {Kanekal}, {Baker}, {Engebretson},
  {Posch}, {Spasojevic}, {Inan}, {Pickett}, and {Decreau}}]{Horne05Nature}
{Horne}, R.~B., {Thorne}, R.~M., {Shprits}, Y.~Y., {Meredith}, N.~P.,
  {Glauert}, S.~A., {Smith}, A.~J., {Kanekal}, S.~G., {Baker}, D.~N.,
  {Engebretson}, M.~J., {Posch}, J.~L., {Spasojevic}, M., {Inan}, U.~S.,
  {Pickett}, J.~S., and {Decreau}, P.~M.~E. (2005{\natexlab{b}}).
\newblock {Wave acceleration of electrons in the Van Allen radiation belts}.
\newblock \emph{Nature}, \textbf{437}, 227--230.

\bibitem[{{Kellogg} \emph{et~al.}(2011){Kellogg}, {Cattell}, {Goetz}, {Monson},
  and {Wilson}}]{Kellogg11}
{Kellogg}, P.~J., {Cattell}, C.~A., {Goetz}, K., {Monson}, S.~J., and {Wilson},
  III, L.~B. (2011).
\newblock {Large amplitude whistlers in the magnetosphere observed with
  Wind-Waves}.
\newblock \emph{\jgr}, \textbf{116}, 9224.

\bibitem[{{Kennel} and {Engelmann}(1966)}]{Kennel&Engelmann66}
{Kennel}, C.~F. and {Engelmann}, F. (1966).
\newblock {Velocity Space Diffusion from Weak Plasma Turbulence in a Magnetic
  Field}.
\newblock \emph{Physics of Fluids}, \textbf{9}, 2377--2388.

\bibitem[{{Kennel} and {Petschek}(1966)}]{Kennel&Petschek66}
{Kennel}, C.~F. and {Petschek}, H.~E. (1966).
\newblock {Limit on Stably Trapped Particle Fluxes}.
\newblock \emph{\jgr}, \textbf{71}, 1--28.

\bibitem[{{Kletzing} \emph{et~al.}(2013){Kletzing}, {Kurth}, {Acuna},
  {MacDowall}, {Torbert}, {Averkamp}, {Bodet}, {Bounds}, {Chutter},
  {Connerney}, {Crawford}, {Dolan}, {Dvorsky}, {Hospodarsky}, {Howard},
  {Jordanova}, {Johnson}, {Kirchner}, {Mokrzycki}, {Needell}, {Odom}, {Mark},
  {Pfaff}, {Phillips}, {Piker}, {Remington}, {Rowland}, {Santolik}, {Schnurr},
  {Sheppard}, {Smith}, {Thorne}, and {Tyler}}]{Kletzing13}
{Kletzing}, C.~A., {Kurth}, W.~S., {Acuna}, M., {MacDowall}, R.~J., {Torbert},
  R.~B., {Averkamp}, T., {Bodet}, D., {Bounds}, S.~R., {Chutter}, M.,
  {Connerney}, J., {Crawford}, D., {Dolan}, J.~S., {Dvorsky}, R.,
  {Hospodarsky}, G.~B., {Howard}, J., {Jordanova}, V., {Johnson}, R.~A.,
  {Kirchner}, D.~L., {Mokrzycki}, B., {Needell}, G., {Odom}, J., {Mark}, D.,
  {Pfaff}, R., {Phillips}, J.~R., {Piker}, C.~W., {Remington}, S.~L.,
  {Rowland}, D., {Santolik}, O., {Schnurr}, R., {Sheppard}, D., {Smith}, C.~W.,
  {Thorne}, R.~M., and {Tyler}, J. (2013).
\newblock {The Electric and Magnetic Field Instrument Suite and Integrated
  Science (EMFISIS) on RBSP}.
\newblock \emph{\ssr}, \textbf{179}, 127--181.

\bibitem[{{Lanzerotti} \emph{et~al.}(1980){Lanzerotti}, {Maclennan},
  {Krimigis}, {Armstrong}, {Behannon}, and {Ness}}]{Lanzerotti80}
{Lanzerotti}, L.~J., {Maclennan}, C.~G., {Krimigis}, S.~M., {Armstrong}, T.~P.,
  {Behannon}, K.~W., and {Ness}, N.~F. (1980).
\newblock {Statics of the nightside Jovian plasma sheet}.
\newblock \emph{\grl}, \textbf{7}, 817--820.

\bibitem[{{Lyons} and {Thorne}(1973)}]{Lyons&Thorne73}
{Lyons}, L.~R. and {Thorne}, R.~M. (1973).
\newblock {Equilibrium structure of radiation belt electrons}.
\newblock \emph{\jgr}, \textbf{78}, 2142--2149.

\bibitem[{{Lyons} \emph{et~al.}(1972){Lyons}, {Thorne}, and {Kennel}}]{Lyons72}
{Lyons}, L.~R., {Thorne}, R.~M., and {Kennel}, C.~F. (1972).
\newblock {Pitch-angle diffusion of radiation belt electrons within the
  plasmasphere.}
\newblock \emph{\jgr}, \textbf{77}, 3455--3474.

\bibitem[{{Malaspina} \emph{et~al.}(2014){Malaspina}, {Andersson}, {Ergun},
  {Wygant}, {Bonnell}, {Kletzing}, {Reeves}, {Skoug}, and
  {Larsen}}]{Malaspina14}
{Malaspina}, D.~M., {Andersson}, L., {Ergun}, R.~E., {Wygant}, J.~R.,
  {Bonnell}, J.~W., {Kletzing}, C., {Reeves}, G.~D., {Skoug}, R.~M., and
  {Larsen}, B.~A. (2014).
\newblock {Nonlinear electric field structures in the inner magnetosphere}.
\newblock \emph{\grl}, \textbf{41}, 5693--5701.

\bibitem[{Malaspina \emph{et~al.}(2015)Malaspina, Wygant, Ergun, Reeves, Skoug,
  and Larsen}]{Malaspina15}
Malaspina, D.~M., Wygant, J.~R., Ergun, R.~E., Reeves, G.~D., Skoug, R.~M., and
  Larsen, B.~A. (2015).
\newblock Electric field structures and waves at plasma boundaries in the inner
  magnetosphere.
\newblock \emph{\jgr}, \textbf{120}, n/a--n/a.
\newblock ISSN 2169-9402.
\newblock 2015JA021137.

\bibitem[{{Mauk} \emph{et~al.}(2013){Mauk}, {Fox}, {Kanekal}, {Kessel},
  {Sibeck}, and {Ukhorskiy}}]{Mauk13}
{Mauk}, B.~H., {Fox}, N.~J., {Kanekal}, S.~G., {Kessel}, R.~L., {Sibeck},
  D.~G., and {Ukhorskiy}, A. (2013).
\newblock {Science Objectives and Rationale for the Radiation Belt Storm Probes
  Mission}.
\newblock \emph{\ssr}, \textbf{179}, 3--27.

\bibitem[{{Meredith} \emph{et~al.}(2004){Meredith}, {Horne}, {Thorne},
  {Summers}, and {Anderson}}]{Meredith_etal2004}
{Meredith}, N.~P., {Horne}, R.~B., {Thorne}, R.~M., {Summers}, D., and
  {Anderson}, R.~R. (2004).
\newblock {Substorm dependence of plasmaspheric hiss}.
\newblock \emph{Journal of Geophysical Research (Space Physics)}, \textbf{109},
  6209.

\bibitem[{{Meredith} \emph{et~al.}(2009){Meredith}, {Horne}, {Glauert},
  {Baker}, {Kanekal}, and {Albert}}]{Meredith09}
{Meredith}, N.~P., {Horne}, R.~B., {Glauert}, S.~A., {Baker}, D.~N., {Kanekal},
  S.~G., and {Albert}, J.~M. (2009).
\newblock {Relativistic electron loss timescales in the slot region}.
\newblock \emph{\jgr}, \textbf{114}, 3222.

\bibitem[{{Mourenas} \emph{et~al.}(2012{\natexlab{a}}){Mourenas}, {Artemyev},
  {Agapitov}, and {Krasnoselskikh}}]{Mourenas12:JGR:acceleration}
{Mourenas}, D., {Artemyev}, A., {Agapitov}, O., and {Krasnoselskikh}, V.
  (2012{\natexlab{a}}).
\newblock {Acceleration of radiation belts electrons by oblique chorus waves}.
\newblock \emph{\jgr}, \textbf{117}, 10\,212.

\bibitem[{{Mourenas} \emph{et~al.}(2012{\natexlab{b}}){Mourenas}, {Artemyev},
  {Ripoll}, {Agapitov}, and {Krasnoselskikh}}]{Mourenas12:JGR}
{Mourenas}, D., {Artemyev}, A.~V., {Ripoll}, J.-F., {Agapitov}, O.~V., and
  {Krasnoselskikh}, V.~V. (2012{\natexlab{b}}).
\newblock {Timescales for electron quasi-linear diffusion by parallel and
  oblique lower-band Chorus waves.}
\newblock \emph{\jgr}, \textbf{117}, A06\,234.

\bibitem[{{Mourenas} \emph{et~al.}(2014){Mourenas}, {Artemyev}, {Agapitov}, and
  {Krasnoselskikh}}]{Mourenas14}
{Mourenas}, D., {Artemyev}, A.~V., {Agapitov}, O.~V., and {Krasnoselskikh}, V.
  (2014).
\newblock {Consequences of geomagnetic activity on energization and loss of
  radiation belt electrons by oblique chorus waves}.
\newblock \emph{\jgr}, \textbf{119}, 2775--2796.

\bibitem[{{Mourenas} \emph{et~al.}(2015){Mourenas}, {Artemyev}, {Agapitov},
  {Krasnoselskikh}, and {Mozer}}]{Mourenas15}
{Mourenas}, D., {Artemyev}, A.~V., {Agapitov}, O.~V., {Krasnoselskikh}, V., and
  {Mozer}, F.~S. (2015).
\newblock {Very oblique whistler generation by low-energy electron streams}.
\newblock \emph{\jgr}, \textbf{120}, 3665–3683.

\bibitem[{{Mozer} \emph{et~al.}(2013){Mozer}, {Bale}, {Bonnell}, {Chaston},
  {Roth}, and {Wygant}}]{Mozer13}
{Mozer}, F.~S., {Bale}, S.~D., {Bonnell}, J.~W., {Chaston}, C.~C., {Roth}, I.,
  and {Wygant}, J. (2013).
\newblock {Megavolt Parallel Potentials Arising from Double-Layer Streams in
  the Earth's Outer Radiation Belt}.
\newblock \emph{Physical Review Letters}, \textbf{111}(23), 235\,002.

\bibitem[{{Mozer} \emph{et~al.}(2014){Mozer}, {Agapitov}, {Krasnoselskikh},
  {Lejosne}, {Reeves}, and {Roth}}]{Mozer14}
{Mozer}, F.~S., {Agapitov}, O., {Krasnoselskikh}, V., {Lejosne}, S., {Reeves},
  G.~D., and {Roth}, I. (2014).
\newblock {Direct Observation of Radiation-Belt Electron Acceleration from
  Electron-Volt Energies to Megavolts by Nonlinear Whistlers}.
\newblock \emph{Physical Review Letters}, \textbf{113}(3), 035\,001.

\bibitem[{{Mozer} \emph{et~al.}(2015){Mozer}, {Agapitov}, {Artemyev}, {Drake},
  {Krasnoselskikh}, {Lejosne}, and {Vasko}}]{Mozer15}
{Mozer}, F.~S., {Agapitov}, O., {Artemyev}, A., {Drake}, J.~F.,
  {Krasnoselskikh}, V., {Lejosne}, S., and {Vasko}, I. (2015).
\newblock {Time domain structures: What and where they are, what they do, and
  how they are made}.
\newblock \emph{\grl}, \textbf{42}, 3627–3638.

\bibitem[{{Omura} \emph{et~al.}(2007){Omura}, {Furuya}, and
  {Summers}}]{Omura07}
{Omura}, Y., {Furuya}, N., and {Summers}, D. (2007).
\newblock {Relativistic turning acceleration of resonant electrons by coherent
  whistler mode waves in a dipole magnetic field}.
\newblock \emph{\jgr}, \textbf{112}, 6236.

\bibitem[{{Reeves} \emph{et~al.}(2013){Reeves}, {Spence}, {Henderson},
  {Morley}, {Friedel}, {Funsten}, {Baker}, {Kanekal}, {Blake}, {Fennell},
  {Claudepierre}, {Thorne}, {Turner}, {Kletzing}, {Kurth}, {Larsen}, and
  {Niehof}}]{Reeves13}
{Reeves}, G.~D., {Spence}, H.~E., {Henderson}, M.~G., {Morley}, S.~K.,
  {Friedel}, R.~H.~W., {Funsten}, H.~O., {Baker}, D.~N., {Kanekal}, S.~G.,
  {Blake}, J.~B., {Fennell}, J.~F., {Claudepierre}, S.~G., {Thorne}, R.~M.,
  {Turner}, D.~L., {Kletzing}, C.~A., {Kurth}, W.~S., {Larsen}, B.~L., and
  {Niehof}, J.~T. (2013).
\newblock {Electron Acceleration in the Heart of the Van Allen Radiation
  Belts}.
\newblock \emph{Science}, \textbf{341}, 991--994.

\bibitem[{{Santol{\'{\i}}k} \emph{et~al.}(2005){Santol{\'{\i}}k}, {Gurnett},
  {Pickett}, {Parrot}, and {Cornilleau-Wehrlin}}]{Santolik05}
{Santol{\'{\i}}k}, O., {Gurnett}, D.~A., {Pickett}, J.~S., {Parrot}, M., and
  {Cornilleau-Wehrlin}, N. (2005).
\newblock {Central position of the source region of storm-time chorus}.
\newblock \emph{\planss}, \textbf{53}, 299--305.

\bibitem[{{Schulz} and {Lanzerotti}(1974)}]{SchulzLanzerotti1974book}
{Schulz}, M. and {Lanzerotti}, L.~J. (1974).
\newblock \emph{{Particle diffusion in the radiation belts}}.

\bibitem[{{Shprits} \emph{et~al.}(2006){Shprits}, {Thorne}, {Horne}, and
  {Summers}}]{Shprits06}
{Shprits}, Y.~Y., {Thorne}, R.~M., {Horne}, R.~B., and {Summers}, D. (2006).
\newblock {Bounce-averaged diffusion coefficients for field-aligned chorus
  waves}.
\newblock \emph{\jgr}, \textbf{111}, 10\,225.

\bibitem[{{Shprits} \emph{et~al.}(2007){Shprits}, {Meredith}, and
  {Thorne}}]{Shprits07}
{Shprits}, Y.~Y., {Meredith}, N.~P., and {Thorne}, R.~M. (2007).
\newblock {Parameterization of radiation belt electron loss timescales due to
  interactions with chorus waves}.
\newblock \emph{\grl}, \textbf{34}, 11\,110.

\bibitem[{{Shprits} \emph{et~al.}(2008{\natexlab{a}}){Shprits}, {Subbotin},
  {Meredith}, and {Elkington}}]{Shprits08:JASTP_local}
{Shprits}, Y.~Y., {Subbotin}, D.~A., {Meredith}, N.~P., and {Elkington}, S.~R.
  (2008{\natexlab{a}}).
\newblock {Review of modeling of losses and sources of relativistic electrons
  in the outer radiation belt II: Local acceleration and loss}.
\newblock \emph{Journal of Atmospheric and Solar-Terrestrial Physics},
  \textbf{70}, 1694--1713.

\bibitem[{{Shprits} \emph{et~al.}(2008{\natexlab{b}}){Shprits}, {Subbotin},
  {Meredith}, and {Elkington}}]{Shprits_etal}
{Shprits}, Y.~Y., {Subbotin}, D.~A., {Meredith}, N.~P., and {Elkington}, S.~R.
  (2008{\natexlab{b}}).
\newblock {Review of modeling of losses and sources of relativistic electrons
  in the outer radiation belt II: Local acceleration and loss}.
\newblock \emph{Journal of Atmospheric and Solar-Terrestrial Physics},
  \textbf{70}, 1694--1713.

\bibitem[{{Spence} \emph{et~al.}(2013){Spence}, {Reeves}, {Baker}, {Blake},
  {Bolton}, {Bourdarie}, {Chan}, {Claudepierre}, {Clemmons}, {Cravens},
  {Elkington}, {Fennell}, {Friedel}, {Funsten}, {Goldstein}, {Green},
  {Guthrie}, {Henderson}, {Horne}, {Hudson}, {Jahn}, {Jordanova}, {Kanekal},
  {Klatt}, {Larsen}, {Li}, {MacDonald}, {Mann}, {Niehof}, {O'Brien}, {Onsager},
  {Salvaggio}, {Skoug}, {Smith}, {Suther}, {Thomsen}, and {Thorne}}]{Spence13}
{Spence}, H.~E., {Reeves}, G.~D., {Baker}, D.~N., {Blake}, J.~B., {Bolton}, M.,
  {Bourdarie}, S., {Chan}, A.~A., {Claudepierre}, S.~G., {Clemmons}, J.~H.,
  {Cravens}, J.~P., {Elkington}, S.~R., {Fennell}, J.~F., {Friedel}, R.~H.~W.,
  {Funsten}, H.~O., {Goldstein}, J., {Green}, J.~C., {Guthrie}, A.,
  {Henderson}, M.~G., {Horne}, R.~B., {Hudson}, M.~K., {Jahn}, J.-M.,
  {Jordanova}, V.~K., {Kanekal}, S.~G., {Klatt}, B.~W., {Larsen}, B.~A., {Li},
  X., {MacDonald}, E.~A., {Mann}, I.~R., {Niehof}, J., {O'Brien}, T.~P.,
  {Onsager}, T.~G., {Salvaggio}, D., {Skoug}, R.~M., {Smith}, S.~S., {Suther},
  L.~L., {Thomsen}, M.~F., and {Thorne}, R.~M. (2013).
\newblock {Science Goals and Overview of the Radiation Belt Storm Probes (RBSP)
  Energetic Particle, Composition, and Thermal Plasma (ECT) Suite on NASA's Van
  Allen Probes Mission}.
\newblock \emph{\ssr}, \textbf{179}, 311--336.

\bibitem[{{Thorne} \emph{et~al.}(1973){Thorne}, {Smith}, {Burton}, and
  {Holzer}}]{1973JGR....78.1581T}
{Thorne}, R.~M., {Smith}, E.~J., {Burton}, R.~K., and {Holzer}, R.~E. (1973).
\newblock {Plasmaspheric Hiss}.
\newblock \emph{\jgr}, \textbf{78}, 1581--1596.

\bibitem[{{Thorne} \emph{et~al.}(2010){Thorne}, {Ni}, {Tao}, {Horne}, and
  {Meredith}}]{Thorne10:Natur}
{Thorne}, R.~M., {Ni}, B., {Tao}, X., {Horne}, R.~B., and {Meredith}, N.~P.
  (2010).
\newblock {Scattering by chorus waves as the dominant cause of diffuse auroral
  precipitation}.
\newblock \emph{\nat}, \textbf{467}, 943--946.

\bibitem[{{Thorne} \emph{et~al.}(2013){Thorne}, {Li}, {Ni}, {Ma}, {Bortnik},
  {Chen}, {Baker}, {Spence}, {Reeves}, {Henderson}, {Kletzing}, {Kurth},
  {Hospodarsky}, {Blake}, {Fennell}, {Claudepierre}, and
  {Kanekal}}]{Thorne13:nature}
{Thorne}, R.~M., {Li}, W., {Ni}, B., {Ma}, Q., {Bortnik}, J., {Chen}, L.,
  {Baker}, D.~N., {Spence}, H.~E., {Reeves}, G.~D., {Henderson}, M.~G.,
  {Kletzing}, C.~A., {Kurth}, W.~S., {Hospodarsky}, G.~B., {Blake}, J.~B.,
  {Fennell}, J.~F., {Claudepierre}, S.~G., and {Kanekal}, S.~G. (2013).
\newblock {Rapid local acceleration of relativistic radiation-belt electrons by
  magnetospheric chorus}.
\newblock \emph{\nat}, \textbf{504}, 411--414.

\bibitem[{{Trakhtengerts}(1966)}]{Trakhtengerts66}
{Trakhtengerts}, V.~Y. (1966).
\newblock {Stationary states of the Earth's outer radiation zone}.
\newblock \emph{Geomagnetism and Aeronomy}, \textbf{6}, 827--836.

\bibitem[{{Trakhtengerts} \emph{et~al.}(2003){Trakhtengerts}, {Rycroft},
  {Nunn}, and {Demekhov}}]{Trakhtengerts03}
{Trakhtengerts}, V.~Y., {Rycroft}, M.~J., {Nunn}, D., and {Demekhov}, A.~G.
  (2003).
\newblock {Cyclotron acceleration of radiation belt electrons by whistlers}.
\newblock \emph{\jgr}, \textbf{108}, 1138.

\bibitem[{{Tsurutani} and {Smith}(1974)}]{Tsurutani&Smith74}
{Tsurutani}, B.~T. and {Smith}, E.~J. (1974).
\newblock {Postmidnight chorus: A substorm phenomenon}.
\newblock \emph{\jgr}, \textbf{79}, 118--127.

\bibitem[{{Vasko} \emph{et~al.}(2015{\natexlab{a}}){Vasko}, {Agapitov},
  {Mozer}, and {Artemyev}}]{Vasko15:jgr:DL}
{Vasko}, I.~Y., {Agapitov}, O.~V., {Mozer}, F., and {Artemyev}, A.~V.
  (2015{\natexlab{a}}).
\newblock {Thermal electron acceleration by electric field spikes in the outer
  radiation belt: generation of field-aligned pitch angle distributions}.
\newblock \emph{\jgr}, page submitted.

\bibitem[{{Vasko} \emph{et~al.}(2015{\natexlab{b}}){Vasko}, {Agapitov},
  {Mozer}, {Artemyev}, and {Jovanovic}}]{Vasko15:grl}
{Vasko}, I.~Y., {Agapitov}, O.~V., {Mozer}, F., {Artemyev}, A.~V., and
  {Jovanovic}, D. (2015{\natexlab{b}}).
\newblock {Magnetic field depression within electron holes}.
\newblock \emph{\grl}, \textbf{42}, 2123--2129.

\bibitem[{Vasko \emph{et~al.}(2015)Vasko, Agapitov, Mozer, and
  Artemyev}]{Vasko2015JGR}
Vasko, I.~Y., Agapitov, O.~V., Mozer, F.~S., and Artemyev, A.~V. (2015).
\newblock Thermal electron acceleration by electric field spikes in the outer
  radiation belt: Generation of field-aligned pitch angle distributions.
\newblock \emph{Journal of Geophysical Research: Space Physics}, pages
  n/a--n/a.
\newblock ISSN 2169-9402.
\newblock \urlprefix\url{http://dx.doi.org/10.1002/2015JA021644}.
\newblock 2015JA021644.

\bibitem[{{Wilson} \emph{et~al.}(2012){Wilson}, {Koval}, {Szabo}, {Breneman},
  {Cattell}, {Goetz}, {Kellogg}, {Kersten}, {Kasper}, {Maruca}, and
  {Pulupa}}]{Wilson12}
{Wilson}, III, L.~B., {Koval}, A., {Szabo}, A., {Breneman}, A., {Cattell},
  C.~A., {Goetz}, K., {Kellogg}, P.~J., {Kersten}, K., {Kasper}, J.~C.,
  {Maruca}, B.~A., and {Pulupa}, M. (2012).
\newblock {Observations of electromagnetic whistler precursors at supercritical
  interplanetary shocks}.
\newblock \emph{\grl}, \textbf{39}, 8109.

\bibitem[{{Wygant} \emph{et~al.}(2013){Wygant}, {Bonnell}, {Goetz}, {Ergun},
  {Mozer}, {Bale}, {Ludlam}, {Turin}, {Harvey}, {Hochmann}, {Harps}, {Dalton},
  {McCauley}, {Rachelson}, {Gordon}, {Donakowski}, {Shultz}, {Smith},
  {Diaz-Aguado}, {Fischer}, {Heavner}, {Berg}, {Malsapina}, {Bolton}, {Hudson},
  {Strangeway}, {Baker}, {Li}, {Albert}, {Foster}, {Chaston}, {Mann},
  {Donovan}, {Cully}, {Cattell}, {Krasnoselskikh}, {Kersten}, {Brenneman}, and
  {Tao}}]{Wygant13}
{Wygant}, J.~R., {Bonnell}, J.~W., {Goetz}, K., {Ergun}, R.~E., {Mozer}, F.~S.,
  {Bale}, S.~D., {Ludlam}, M., {Turin}, P., {Harvey}, P.~R., {Hochmann}, R.,
  {Harps}, K., {Dalton}, G., {McCauley}, J., {Rachelson}, W., {Gordon}, D.,
  {Donakowski}, B., {Shultz}, C., {Smith}, C., {Diaz-Aguado}, M., {Fischer},
  J., {Heavner}, S., {Berg}, P., {Malsapina}, D.~M., {Bolton}, M.~K., {Hudson},
  M., {Strangeway}, R.~J., {Baker}, D.~N., {Li}, X., {Albert}, J., {Foster},
  J.~C., {Chaston}, C.~C., {Mann}, I., {Donovan}, E., {Cully}, C.~M.,
  {Cattell}, C.~A., {Krasnoselskikh}, V., {Kersten}, K., {Brenneman}, A., and
  {Tao}, J.~B. (2013).
\newblock {The Electric Field and Waves Instruments on the Radiation Belt Storm
  Probes Mission}.
\newblock \emph{\ssr}, \textbf{179}, 183--220.

\end{thebibliography}

\end{document}